\documentclass[a4paper,11pt,bibtex]{article}
\usepackage{pos}

\usepackage{mathptmx}  
\usepackage{amsmath}  
\usepackage{amssymb} 
\usepackage{siunitx}
\usepackage{natbibspacing}

\def\mnras{, MNRAS}
\def\apj{, ApJ}
\def\apjs{, ApJS}
\def\aap{, A\&A}
\def\aj{, AJ}
\def\apss{, astroph. \& space science}
\def\aapr{, A\&A reviews}
\def\deg{$^\circ$}
\def\arcmin{$^\prime$}
\def\arcsec{$^{\prime\prime}$}

\title{LMC N132D: a mature supernova remnant with a youthful gamma-ray spectrum}

 \ShortTitle{LMC N132D: a mature SNR with a youthful spectrum}

\author*[a]{Jacco Vink}
\author[a]{Rachel Simoni}
\author[b]{Nukri Komin}
\author[a]{Dmitry Prokhorov}

\affiliation[a]{University of Amsterdam, Anton Pannekoek Institute \& GRAPPA,\\ 
  Science Park 904, 1098 XH, Amsterdam, The Netherlands}

\affiliation[b]{
University of the Witwatersrand, School of Physics,\\
1 Jan Smuts Avenue, Braamfontein, Johannesburg, 2050 South Africa
}

\forColl{H.E.S.S.} 

\emailAdd{j.vink@uva.nl}
\emailAdd{r.simoni@uva.nl}
\emailAdd{nukri.komin@wits.ac.za}
\emailAdd{d.prokhorov@uva.nl}

\abstract{ 
The supernova remnant LMC N132D is a remarkably luminous gamma-ray emitter at $\sim$50 kpc with an age of $\sim$2500 years. It belongs to the small group of oxygen-rich SNRs, which includes Cassiopeia A (Cas A) and Puppis A. N132D is interacting with a nearby molecular cloud. By adding 102 hours of new observations with the High Energy Stereoscopic System (H.E.S.S.) to the previously published data with exposure time of 150 hours, we achieve the significant detection of N132D at a 5.7$\sigma$ level in the very high energy (VHE) domain. The gamma-ray spectrum is compatible with a single power law extending above 10 TeV. We set a lower limit on an exponential cutoff energy at 8 TeV with 95\% CL. The multi-wavelength study supports a hadronic origin of VHE gamma-ray emission indicating the presence of sub-PeV cosmic-ray protons. The detection of N132D is remarkable since the TeV luminosity is higher than that of Cas A by more than an order of magnitude. Its luminosity is comparable to, or even exceeding the luminosity of RX J1713.7-3946 or HESS J1640-465. Moreover, the extended power-law tail in the VHE spectrum of N132D is surprising given both the exponential cutoff at 3.5 TeV in the spectrum of its 340-year-old sibling, Cassiopeia A, and the lack of TeV emission from a Fermi-LAT 2FHL source ($E>50$~GeV) associated with Puppis A. We discuss a physical scenario leading to the enhancement of TeV emission via the interaction between N132D and a near molecular cloud.
}

\FullConference{37$^{\rm{th}}$ International Cosmic Ray Conference (ICRC 2021)\\
		July 12th -- 23rd, 2021\\
		Online -- Berlin, Germany}

\begin{document}
\maketitle

\section{Introduction}

Since the 1960s 
 supernova remnants (SNRs) are suspected to be the dominant contributors to the Galactic cosmic-ray spectrum  \citep{ginzburg64b}.
The main arguments are that the Galactic cosmic-ray energy budget fits well with the power provided by supernovae, if 10\% of the explosion energy is used for
accelerating cosmic rays. Moreover, SNRs are prominent radio synchrotron sources, indicating that SNRs contain ample amounts of electron (leptonic) cosmic rays.
The cosmic-ray spectrum is nearly featureless from $\sim 10$~GeV up to $\sim 3$~PeV, which implies that if SNRs are the dominant
sources of cosmic rays, and they must be able to accelerate protons to  $\sim 3$~PeV and more heavy atomic nuclei to even higher energies 
(hadronic cosmic rays).

Some two decades ago  the hope was that gamma-ray observation would be able to provide evidence that SNRs are indeed the sources of hadronic
cosmic rays. Despite the wealth of information gamma-ray observations have since then provided, 
it is still uncertain whether SNRs are indeed the dominant sources of  cosmic rays, and capable of accelerating protons up to 3~PeV.
Broad-band spectral energy distributions (SEDs) of some SNRs
have provided clear evidence for the presence of hadronic cosmic rays in several SNRs. 
But many of the more mature  SNRs have gamma-ray spectra with broken power-law distributions, indicating that the highest energy cosmic rays have already
diffused away from the SNR shells. Examples are IC 443 and W44, whose spectra clearly display the "pion bump" feature, but also have spectral
breaks  indicating that a large fraction of the protons above energies of $\sim 20$--$200$~GeV have escaped \citep{ackermann13}. 
The situation for young SNRs is even more puzzling.
Among the young SNRs, the two best candidates for  hadronic gamma-ray emission are Tycho's SNR \citep{veritas17_tycho} and Cassiopeia A (Cas A) \citep{magic17_casa,veritas20_casa}, and for both SNRs the gamma-ray spectra indicate cutoff energies of 1.7~TeV ({ but at $1.8\sigma$ level}) and 3.5~TeV, respectively. 

Our understanding of the cosmic-ray spectrum itself is evolving as well:
the different spectral shape of the helium and proton cosmic rays suggests multiple origins for the overal cosmic-ray spectrum \citep{pamela11},
and there is evidence
for spectral steepening of the proton spectrum starting  around 10~TeV \citep{lipari20}. Both features  suggest that multiple populations
of cosmic-ray sources may contribute to the cosmic-ray spectrum --- either very different types of sources, or subclasses of supernovae/SNRs
 \citep[c.f.][]{ptuskin03,schure13}. Those supernovae (SNe)  that produce multi-PeV protons
may do so shortly after the SN explosion, provided  the SN is surrounded by a dense stellar wind \citep[e.g.][]{cardillo15,marcowith18,hess_supernovae19}. 

\begin{figure}
\begin{minipage}[l]{0.6\textwidth}
\centerline{\includegraphics[width=0.9\textwidth]{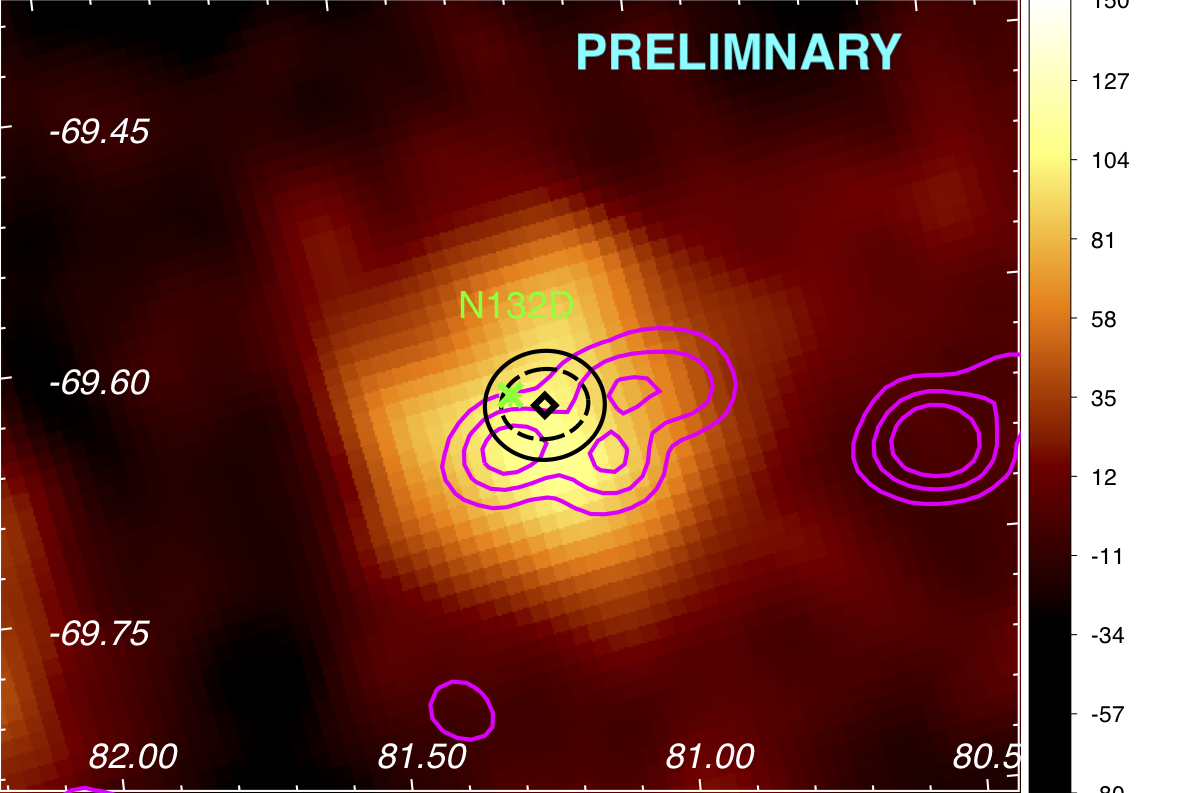}}
\end{minipage}
\begin{minipage}[l]{0.4\textwidth}
\caption{\label{fig:excess}
H.E.S.S. excess events map of a 1.6\deg by 1.6\deg\ region centered on N132D. The map has been smoothed with Gaussian with a kernel width of $\sigma=0.1$\deg.
The position of N132D is indicated by the diamond, and the best fit position is indicated by the red $2\sigma$ and $3\sigma$ contours.
In magenta  $J=1\rightarrow 0$ line emission  from carbon-monoxide (CO) is indicated, as obtained by the MAGMA survey \citep{magma11}
}
\end{minipage}
\end{figure}

In light of these discussions the Large Magellanic Cloud (LMC) SNR N132D is a very interesting gamma-ray source. 
Here we present updated gamma-ray observations by H.E.S.S. of this luminous SNR \citep{hess_lmc}, together with data obtained by the Fermi-LAT instrument.
With an estimated age of $\sim 2500$~yr \citep{vogt11}, N132D is considerably older than Cas A and Tycho's SNR ($\sim$340 yr and 449~yr respectively).
We report here that N132D has, despite its age, a gamma-ray spectrum that can be well fit by a single power law extending beyond 10~TeV.

N132D is labeled as an "oxygen-rich SNR", a category that includes Cas A and Puppis A, but the latter is not a luminous TeV gamma-ray source \citep{hess_puppisa}. N132D is among the most luminous SNRs in the radio domain \citep{dickel95}.
The SNR has a radius of approximately 0.8\arcmin, corresponding to 3.7~pc at the distance of the LMC (50~kpc). 
X-ray measurements suggest that N132D was created by a SN  with an unusually large  explosion energy of $6\times 10^{51}$~erg \citep{hughes98}.
In \cite{hughes98} it was argued that the SNR has been  evolving in a wind-blown cavity, and that the shock
has  relatively recently started interacting with the surrounding wind shell. The SNR, or its progenitor star, has carved out a hole on the northeastern edge of an extended 
HI cloud, in which  several dense molecular cloud are present southwest of the SNR  \citep{sano20}.  
The presence of enhanced  CO $J=3\rightarrow 2$ over $J=1\rightarrow 0$ emission at southern edge of the SNR as well as toward the center of the SNR shows
that the SNR shock is interacting with molecular gas in some locations of the SNR \citep{sano20}.

N132D was first detected by H.E.S.S. at the 4.7$\sigma$ level (pre-trial) \citep{hess_lmc}. 
Additional observations by H.E.S.S. have now increased the significance to $5.7\sigma$. 
These proceedings summarise the results of a H.E.S.S. collaboration publication recently accepted by Astronomy \& Astrophysics \citep{hess_n132d_2021}.

\section{Observations}
The measurements presented here were made with H.E.S.S., an imaging atmospheric Cherenkov telescope array of four 12~m telescopes (CT1--4) and one 28~m telescope (CT5),
located in the Khomas Highland of Namibia. 
H.E.S.S. observed the  LMC region containing N132D for 252 hours using the CT1--4  telescopes.
Apart from N132D, this region also contains
the Doradus region, which contains two other gamma-ray sources, the pulsar wind nebula N157D and the superbubble 30DorC \citep{hess_lmc}.
Also SN1987A is located in this region, but has not yet been detected.

As the LMC is  observed at large zenith angles, the energy threshold is rather high, and we have used only gamma-ray events with reconstructed
energies above 1.3~TeV. As is common practice with the H.E.S.S. collaboration the data were analysed by two independent analysis chains. Here we present
the results obtained with the ImPACT algorithm for event reconstruction \citep{parsons14}. We produced excess maps  using the
ring background method, whereas for spectral extraction we used the reflected background method \citep{berge07}.

Fig.~\ref{fig:excess}  shows the resulting event excess map, clearing indicating a large excess coinciding with the position of N132D. The nominal position of
the gamma-ray source, obtained assuming a point source convolved with the H.E.S.S. point spread function gives
RA and DEC as  5h24m47s ($\pm$6.9s) and DEC -69\deg38\arcmin50\arcsec\ ($\pm$29\arcsec), which is about 1\arcmin\  from the X-ray source position.
But the gamma-ray position is consistent with the center of N132D at the $2\sigma$ level.

\begin{figure}
\centerline{\includegraphics[angle=90,trim=0 0 0 0,clip=true,width=0.5\textwidth]{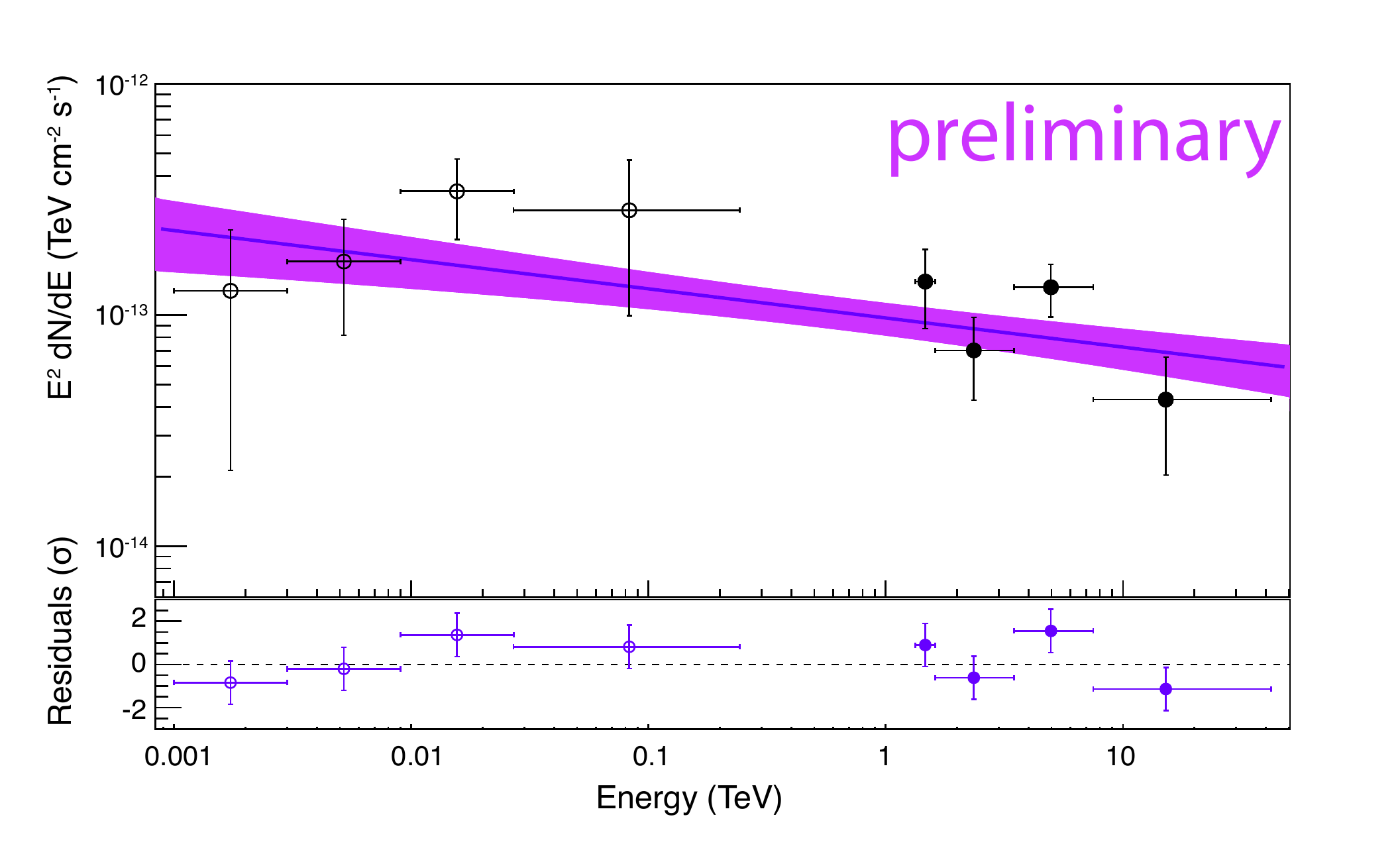}
\includegraphics[width=0.5\textwidth]{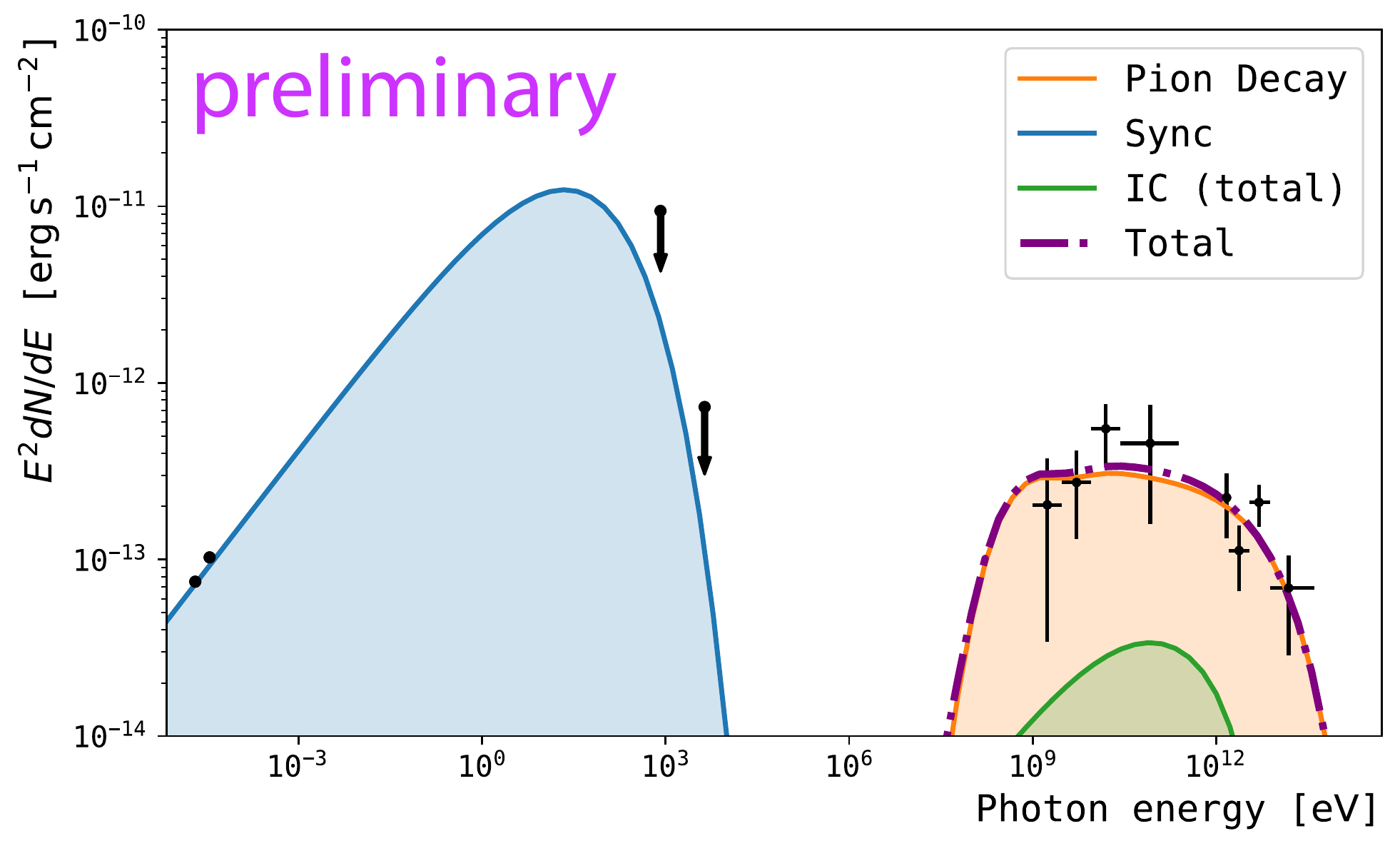}
}
\caption{\label{fig:spectrum}
Left panel:
H.E.S.S. (solid circles) and Fermi-LAT (open circles) spectral energy density points.
The red line and shaded area shows the best fit unbroken power-law fit to the combined data points and its uncertainty, respectively. 
Right panel: Broad-band SED with  a model for which the gamma-ray emission is dominated by pion production and decay---i.e. a hadronic model.
The radio data points are taken from \cite{dickel95}, and the upper limits on non-thermal X-ray emission from \cite{bamba18}.
}
\end{figure}

Fig.~\ref{fig:spectrum} shows the spectral energy distribution augmented
with the spectral data of the Fermi-LAT \citep{atwood09} experiment, based on 10.8 yr of observations, and using the
FERMITOOLS v1.0.7 software package for modeling the maps and performing spectral extractions, using the
P8R3\_SOURCE\_V2  instrument data.

The broad gamma-ray spectrum is well fit with a single power-law distribution with spectral index $\Gamma=2.13\pm 0.05$, and  a flux normalisation at
1~TeV of  $(9.7 \pm 1.6) \times 10^{-14}$~TeV$^{-1}$cm$^{-2}$s$^{-1}$ (Fig.~\ref{fig:spectrum}).
The fit statistic is $\chi^2/d.o.f =8.2/6$.  We also tested a broken power-law spectrum and found a best fit for the power-law indices of $\Gamma_1=1.47\pm 0.43$ and $\Gamma_1=2.31\pm 0.14$,
with a break energy of $E_b=24^{116}_{-16}$~GeV.
The fit statistic of  $\chi^2/d.o.f =4.2/4$ indicates that the improvement is only marginal ($\Delta \chi^2\approx 2$ per additional degree of freedom).
Therefore, a single power-law is to be preferred, based on the current data sets.

Finally, we  tested for power-law spectrum with exponential cutoff, a model that best fits the Cas A very-high energy spectrum.
Also in this case a single power-law in preferred. Assuming that there
is an exponential cutoff, we constrained the cutoff to be $E_{\rm cut}=19^{+60}_{-10}$~TeV. Also in this case
the statistical improvement over a single power-law distribution is marginal, with $\chi^2/d.o.f.=6.4/5$. Moreover, the best fit value for $E_{\rm cut}$ is larger
than the fitted energy range.

The observed spectrum itself indicates that N132D is indeed a very luminous SNR in gamma rays. The inferred luminosity of $L(1-10~{\rm TeV})= (1.05\pm0.03)\times
10^{35}$~erg\ s$^{-1}$ is comparable to HESS J1640-465 and $\sim 28$ times more luminous than Cas A.

\section{Discussion}

The combined H.E.S.S. and Fermi-LAT data discussed here show that also in gamma-rays N132D has some remarkable properties.
In order to understand these better we relied on physically modelling the radio to gamma-ray SED.

\subsection{The nature of the gamma-ray emission mechanism}
First of all, the gamma-ray spectral index for the joint H.E.S.S. and Fermi-LAT spectral points of $\Gamma=2.2$ suggests already that most likely
the gamma-ray emission has a hadronic origin. For a leptonic model, dominated by inverse Compton scattering, the  photon spectral  index is related
to the radio spectral energy density index $\alpha$ as $\Gamma=\alpha +1$. The radio spectral index of $\alpha=0.7$ \citep{dickel95} of N132D therefore predicts
an inverse Compton scattering photon index of $\Gamma=1.7$. Moreover, it is worth noting that the X-ray observations of N132D  do not provide
evidence for X-ray synchrotron emission \citep{bamba18}, unlike younger, TeV gamma-ray emitting SNRs like Cas A,  Kepler's, Tycho's SNR, SN 1006, etc.
 \citep[e.g.][for a review]{vink12}, or even 30DorC, which is also located in the LMC \citep{hess_lmc,kavanagh19}.

For a hadronic origin of the gamma-ray emission---i.e. pion decay---the photon index
corresponds roughly with  the spectral index of the primary cosmic-ray spectrum. The measured photon index of $\Gamma=2.3$ is comparable to those
of other SNRs with hadronic gamma-ray emission, as well as the expected spectral index for Galactic cosmic-ray sources.

A more detailed modelling\footnote{We used the Naima package \citep{naima} to perform both the leptonic and hadronic  modeling. For the leptonic model
we toke into account both the cosmic-microwave background and local infrared radiation energy density.}
 of the SED of N132D with a leptonic model shows that matching both the radio synchrotron flux points with the gamma-ray
data points requires a relatively low magnetic field strength of $B\approx 20~{\rm \mu G}$ and a total electron cosmic-ray energy of $W_{\rm e}\approx 5\times 10^{49}$~erg.
This requires a very high conversion efficiency of explosion energy to electron cosmic-ray energy of just $\sim 2$\%. Another way to look at it is that
if the flux ratio between proton- and electron cosmic rays observed at Earth of $n_{\rm p}/n_{\rm e}\approx 100$ holds also for N132D, the total
cosmic-ray energy in N132D would be $10^{52}$~erg, much larger than the inferred explosion energy of $5\times 10^{51}$~erg.

Given the problems of the leptonic model for explaining the  gamma-ray emission from N132D, a hadronic model seems much more plausible---a conclusion also reached by 
\cite{bamba18}. 
In Fig.~\ref{fig:spectrum}(right)  we show such a hadronic SED model.
The total hadronic cosmic-ray energy for this model is $W_{\rm p}=4\times 10^{50} (n_{\rm p}/10~{\rm cm}^{-3})^{-1}$~erg, with $n_{\rm p}$ the local plasma density in the source.
This value for $W_{\rm p}$ corresponds to roughly 10\% of the available explosion energy, with fits well with the general requirement that SNRs should have
an explosion energy to cosmic-ray energy of 5--10\% in order to explain the cosmic-ray energy budget of the Milky Way.

For the  hadronic model the average magnetic-field strength is a free parameter, provided that the strength is
substantially larger than the best fit value for the leptonic model---i.e. $B\gg 20~{\rm \mu G}$. 
In the model shown in Fig.~\ref{fig:spectrum} $B=100~{\rm \mu G}$, which corresponds to a hadronic over leptonic cosmic-ray energy ratio of
$W_{\rm p}/W_{\rm e}=100 (n_{\rm p}/10~{\rm cm}^{-3})^{-1}$.

\subsection{Possible interactions with local molecular clouds}

N132D is located near a molecular cloud complex, and it has recently been shown that the SNR is interacting with some of the molecular gas \citep{sano20}.
For the hadronic model, preferred by our analysis, the local gas density has to be high $n_{\rm p}\gtrsim 10$~cm$^{-3}$ in order to explain the gamma-ray luminosity
of N132D. The analysis  of observations of HI and CO line emission by \cite{sano20} shows that the density requirements are easily met. The HI gas cloud around
the SNR has a typical density of $n_{\rm H}\approx 30$~cm$^{-3}$, implying $W_{\rm p}\approx 1\times 10^{50}$~erg. The HI cloud has, however, a positive gradient
from the northeast to the southwest surrounding N132D. 

The molecular clouds, in general, are located further away from N132D (Fig.~\ref{fig:excess}), but  a few smaller, shocked clouds are present at the southern edge N132D and at the center of the remnant, with 
 estimated densities ranging from 100~cm$^{-3}$ to 900~cm$^{-3}$ \citep{sano20}.
Sano et al. \cite{sano20} argue that the shocked cloudlets toward the center of N132D are also physically located within the SNR shell, and have survived shock
passage due to their high density.

The question is whether the gamma-ray emission comes from the shell of SNR itself, which is mainly interacting with the dense
10---30~cm$^{-3}$ HI gas, or from a few of the small dense clouds described above. Yet another possibility is that the gamma-ray emission
originates 
from cosmic rays that have escaped the shell and are now engulfing the larger CO clouds toward the southwest of the SNR.
A hint in favor of the latter scenario is that the gamma-ray emission appears not be centered on N132D, but about an arcminute  southwest of the SNR.
Such a scenario would be reminiscent of the situation around the older SNR W28  \citep{hessw28}. However, we stress that the offset between N132D and the gamma-ray source is statistically not significant. 

Which of the above scenarios is correct, is important given that the broad gamma-ray SED is best fit with a single power law with no hint of an exponential cutoff.
If the gamma-ray emission is from the shell of N132D, the SED implies that despite its age of $\sim 2500$~yr, N132D has been able to accelerate cosmic rays at least
up to $\sim 100$~TeV,  and been able to retain these. In contrast, if the gamma-ray emission comes from the molecular clouds in the vicinity of N132D, ongoing
cosmic-ray acceleration in N132D is probably low, and the gamma-ray emission gives us a window on the cosmic-ray acceleration history of N132D.

The fact that a single power-law distribution seems to fit the broad gamma-ray SED best, is in favour of the scenario in which the emission is dominated by 
the SNR itself, because escape of cosmic rays is an energy-dependent process. So one expects in that case that the low-energy gamma-ray emission must
come from the SNR itself, and the higher energy gamma-rays from further away from the SNR. It is difficult in that situation to not have a more complex spectral
distribution than a single power-law distribution \citep{gabici09}.

For now the most likely scenario appears to be one in which N132D seems to be a an efficient accelerator and 
is able to either still be able to accelerate particles up to at least 100~TeV, or has been able to retain these very high energy cosmic rays. 
One could speculate that the evolution in a cavity may have delayed the slowing down of the shock velocity to later ages, resulting in a rather
late conversion of explosion energy to cosmic ray energy.
The lack of a clear spectral cutoff below 10~TeV is in  contrasts to the situation for 
N132D's Galactic counterpart, Cas A, which shows a cutoff in the primary cosmic-ray spectrum around 10~TeV:
when it comes to  its gamma-ray spectrum, the 2500~yr old N132D appears more youthful than its 340~yr old Galactic cousin.

\section{Conclusions}
We reported here that the LMC SNR N132D is now detected by with H.E.S.S. at the 5.7$\sigma$ level. A joint analysis
of the H.E.S.S. and Fermi-LAT spectra shows that the gamma-ray spectrum is well-fitted with a single power law distribution with
photon index $\Gamma=2.13\pm 0.05$. The centroid of gamma-ray emission is located $\sim 1$~arcmin southwest of N132D,
but is statistically consistent with the location of the  N132D SNR. 
The SED is best fit with a hadronic emission model, with a total implied cosmic-ray energy content of 
$W_{\rm p}=4\times 10^{50} (n_{\rm p}/10~{\rm cm}^{-3})^{-1}$~erg.
The lack of a clear cutoff in the spectrum below 10~TeV is of interest, given that the much younger Galactic  SNR Cas A,
to which N132D is often compared, has a cutoff around 3.5 TeV.

It is possible that all, or some, of the gamma-ray emission is not coming from the shell of N132D, but from nearby
molecular clouds. However,
it will require CTA \cite{2019CTAbook}, with its better angular resolution and much higher sensitivity,
to pinpoint the location of the gamma-ray emission, and improve the determination of the spectral shape,
both of which will provide new insights into the cosmic-ray acceleration properties of this very powerful SNR.

\bibliographystyle{JHEP}

\providecommand{\href}[2]{#2}\begingroup\raggedright\endgroup

\

{\scriptsize
\noindent{ Acknowledgements}
The support of the Namibian authorities and of the University of Namibia in facilitating the construction and operation of H.E.S.S. is gratefully acknowledged, as is the support by the German Ministry for Education and Research (BMBF), the Max Planck Society, the German Research Foundation (DFG), the Helmholtz Association, the Alexander von Humboldt Foundation, the French Ministry of Higher Education, Research and Innovation, the Centre National de la Recherche Scientifique (CNRS/IN2P3 and CNRS/INSU), the Commissariat \`a l'\'energie atomique et aux \'energies alternatives (CEA), the U.K. Science and Technology Facilities Council (STFC), the Knut and Alice Wallenberg Foundation, the National Science Centre, Poland grant no. 2016/22/M/ST9/00382, the South African Department of Science and Technology and National Research Foundation, the University of Namibia, the National Commission on Research, Science \& Technology of Namibia (NCRST), the Austrian Federal Ministry of Education, Science and Research and the Austrian Science Fund (FWF), the Australian Research Council (ARC), the Japan Society for the Promotion of Science and by the University of Amsterdam.

We appreciate the excellent work of the technical support staff in Berlin, Zeuthen, Heidelberg, Palaiseau, Paris, Saclay, T\"ubingen and in Namibia in the construction and operation of the equipment. This work benefitted from services provided by the H.E.S.S. Virtual Organisation, supported by the national resource providers of the EGI Federation.
}
\vfill
\pagebreak
\section*{Full Authors List: H.E.S.S. Collaboration}

\scriptsize
\noindent
H.~Abdalla$^{1}$, 
F.~Aharonian$^{2,3,4}$, 
F.~Ait~Benkhali$^{3}$, 
E.O.~Ang\"uner$^{5}$, 
C.~Arcaro$^{6}$, 
C.~Armand$^{7}$, 
T.~Armstrong$^{8}$, 
H.~Ashkar$^{9}$, 
M.~Backes$^{1,6}$, 
V.~Baghmanyan$^{10}$
% ...

\medskip

\noindent
$^{1}$University of Namibia, Department of Physics, Private Bag 13301, Windhoek 10005, Namibia\\
$^{2}$Dublin Institute for Advanced Studies, 31 Fitzwilliam Place, Dublin 2, Ireland\\
$^{3}$Max-Planck-Institut f\"ur Kernphysik, P.O. Box 103980, D 69029 Heidelberg, Germany\\
$^{4}$High Energy Astrophysics Laboratory, RAU,  123 Hovsep Emin St  Yerevan 0051, Armenia\\
$^{5}$Aix Marseille Universit\'e, CNRS/IN2P3, CPPM, Marseille, France\\
$^{6}$Centre for Space Research, North-West University, Potchefstroom 2520, South Africa\\
$^{7}$Laboratoire d'Annecy de Physique des Particules, Univ. Grenoble Alpes, Univ. Savoie Mont Blanc, CNRS, LAPP, 74000 Annecy, France\\
$^{8}$University of Oxford, Department of Physics, Denys Wilkinson Building, Keble Road, Oxford OX1 3RH, UK\\
$^{9}$IRFU, CEA, Universit\'e Paris-Saclay, F-91191 Gif-sur-Yvette, France\\
$^{10}$Instytut Fizyki J\c{a}drowej PAN, ul. Radzikowskiego 152, 31-342 Krak{\'o}w, Poland\\
% ...

\end{document}